\documentclass[doublecol]{epl2}
\usepackage{graphicx}
\usepackage{amsmath}
\usepackage{amsfonts}
\usepackage{amssymb}

\newcommand{\ket}[1]{|#1\rangle}
\newcommand{\bra}[1]{\langle #1|}

\title{Non-Abelian off-diagonal geometric phases in nano-engineered four-qubit systems}
\author{Vahid Azimi Mousolou \inst{1} \and Carlo M. Canali \inst{1} \and 
Erik Sj\"oqvist \inst{2} \inst{3}} 

\institute{\inst{1} Department of Physics and Electrical Engineering, Linnaeus University,
 391 82 Kalmar, Sweden. \\
\inst{2} Department of Quantum Chemistry, Uppsala University,
Box 518, Se-751 20 Uppsala, Sweden. \\
\inst{3} Centre for Quantum Technologies, National University of Singapore,
3 Science Drive 2, 117543 Singapore, Singapore}  


\date{\today}

\abstract{The concept of off-diagonal geometric phase (GP) has been introduced in order 
to recover interference information about the geometry of quantal evolution where the 
standard GPs are not well-defined. In this Letter, we propose a physical setting for realizing 
non-Abelian off-diagonal GPs. The proposed non-Abelian off-diagonal GPs can be 
implemented in a cyclic chain of four qubits with controllable nearest-neighbor interactions.
Our proposal seems to be within reach in various nano-engineered systems and therefore 
opens up for first experimental test of the non-Abelian off-diagonal GP.}
\begin{document}

\maketitle

\section{Introduction}
When a state of a quantal system evolves in time, it may pick up a geometric phase (GP) factor that 
reflects the geometry of the underlying state space. This phase factor turns out to be undefined 
if the end-points of the path correspond to distinguishable, i.e., orthogonal, states. 
This fact led Manini and Pistolesi \cite{manini00} to introduce off-diagonal GPs that capture 
interference information related to state space geometry in cases where the standard GP is not 
defined. This off-diagonal GP was subsequently verified experimentally for neutron spin 
\cite{hasegawa01,hasegawa02} and its mixed state counterpart was identified in Refs. 
\cite{filipp03a,filipp03b,tong05}. 

Kult {\it et al.} \cite{kult07} generalized the off-diagonal GP in Ref. \cite{manini00} to the 
non-Abelian case. A non-Abelian GP is a unitary matrix that reflects the geometry of a 
Grassmann manifold, i.e., a space of subspaces of some given dimension of a complex 
vector space \cite{bengtsson06}. Each subspace along a path in a Grassmann manifold 
may represent the post-measurement state of an incomplete projective measurements 
\cite{anandan89,sjoqvist06} or an encoding of a quantum computational system 
\cite{zanardi99,sjoqvist12}. The non-Abelian setting offers the additional possibility 
of partially overlapping subspaces giving rise to a richer off-diagonal GP structure than 
in the Abelian case. 

Here, we provide an explicit physical setting for the non-Abelian off-diagonal GPs in terms 
of a cyclic chain of four qubits with nearest-neighbor interaction. The setup can be used 
for realizations of non-Abelian off-diagonal GPs in different kinds of nano-engineered 
systems, such as in quantum dots \cite{mousolou12}, atoms in optical lattices \cite{radic12}, and 
topological insulators \cite{zhu11}. Our proposal seems to be within reach with current technology 
and therefore opens up for first experimental test of the non-Abelian off-diagonal GP. 

\section{Non-Abelian off-diagonal GPs}
We first briefly review the basic theory of non-Abelian off-diagonal GPs. Suppose $\mathcal{H}$ 
is the system's Hilbert space and consider the smoothly parametrized decomposition 
\begin{eqnarray}
\mathcal{H} = \mathcal{H}_1 (s) \oplus \cdots \oplus \mathcal{H}_{\eta} (s), \ s \in [0,t]
\end{eqnarray}
into $\eta$ mutually orthogonal subspaces. We assume that $\dim \mathcal{H}_l (s) = n_l$ are 
constant for the duration of the evolution and $\sum_{l=1}^{\eta} n_l = \dim \mathcal{H} \equiv N$. 
The evolution $s \mapsto \mathcal{H}_l (s)$ is a path $\mathcal{C}_l$ in the Grassmann manifold 
$\mathcal{G} (N;n_l)$, i.e., the space of $n_l$-dimensional subspaces of the $N$-dimensional 
Hilbert space $\mathcal{H}$. 

Let $\mathcal{F}_l (s) = \{ l^p (s) \}_{p=1}^{n_l}$ be a smoothly parametrized frame (ordered 
orthonormal basis) spanning the subspace $\mathcal{H}_l (s)$. Define the quantities 
\begin{eqnarray}
\sigma_{kl} = \left[ \mathcal{F}_k (0) | \mathcal{F}_l (t) \right] {\bf T} 
e^{\int_0^t A_l (s) ds} , 
\end{eqnarray}
where $\left[ \mathcal{F}_k (0) | \mathcal{F}_l (t) \right]_{pq} = \langle k^p (0) \ket{l^q (t)}$ 
is the $n_k \times n_l$ overlap matrix 
of the frames $\mathcal{F}_k (0)$ and $\mathcal{F}_l (t)$ and $[ A_l (s)]_{pq} = \langle 
\partial_s l^p (s) \ket{l^q (s)}$ is the Wilczek-Zee connection \cite{wilczek84} along the path 
$\mathcal{C}_l$. 

In the case where $k=l$, the quantity  
\begin{eqnarray} 
U_g^{(1)} (\mathcal{C}_l) = \Phi [\sigma_{ll}] = \left( 
\sqrt{\sigma_{ll} \sigma_{ll}^{\dagger} } \right)^{\ominus} 
\sigma_{ll}, 
\label{eq:diagonal}
\end{eqnarray}
$\ominus$ being the Moore-Penrose pseudoinverse \cite{moore20,penrose55} obtained 
by inverting all nonzero eigenvalues, is the standard open path 
non-Abelian GP \cite{mostafazadeh99,kult06}. If $\sigma_{ll}$ is full rank, 
$\Phi [\sigma_{ll}]$ is a unitary $n_l \times n_l$ matrix; if the rank is nonzero and lower than $n_l$, 
the non-Abelian GP is partially defined \cite{kult06} corresponding to the case where the subspaces 
at the end-points of the path $\mathcal{C}_l$ in the Grassmann manifold $\mathcal{G} (N;n_l)$ 
are partially overlapping, i.e., their nonzero overlap matrix is not full rank. Note that a partially defined $U_g^{(1)} (\mathcal{C}_l)$ is a partial isometry, i.e., an operator such that $U_g^{(1)} (\mathcal{C}_l) 
U_g^{(1)\dagger} (\mathcal{C}_l)$ and $U_g^{(1)\dagger} (\mathcal{C}_l) U_g^{(1)} (\mathcal{C}_l)$ 
are projectors onto the two end-points \cite{kult06}. In the case of orthogonal subspaces at the 
end-points of the path $\mathcal{C}_l$, $\sigma_{ll}$ vanishes and the non-Abelian GP is undefined. 
The points along the evolution, where the GP is undefined or partially defined, are considered as 
singular points of the evolution.

For a set $\{ l_1,\ldots,l_{\kappa} \}$ of distinct indices, each $\sigma_{kl}$, $k,l \in 
\{ l_1,\ldots,l_{\kappa} \}$, transforms as $\sigma_{kl} \rightarrow U_{k}^{\dagger} (0) 
\sigma_{kl} U_{l} (0)$, where the two unitaries $U_{k}^{\dagger} (0)$ and $U_{l}^{\dagger} (0)$ 
induce change of frames of the two subspaces $\mathcal{H}_{k} (0)$ and $\mathcal{H}_{l} (0)$. 
Thus, $\sigma_{kl}$ transforms non-covariantly since $U_k (0)$ and $U_l (0)$ can be chosen independently.  This transformation property suggests 
\begin{eqnarray}
\gamma_{l_1 \ldots l_{\kappa}} = \sigma_{l_1 l_{\kappa}} 
\sigma_{l_{\kappa} l_{\kappa -1}} \cdots \sigma_{l_2 l_1} 
\end{eqnarray}
as gauge covariant quantities in terms of which the non-Abelian off-diagonal GPs of 
order $\kappa$ are defined to be 
\begin{eqnarray}
U_g^{(\kappa)} (\mathcal{C}_{l_1},\ldots , \mathcal{C}_{l_{\kappa}}) & = &  
\Phi \left[ \gamma_{l_1 \ldots l_{\kappa}}\right] 
\nonumber \\ 
 & = & \left( \sqrt{ \gamma_{l_1 \ldots l_{\kappa}} \gamma_{l_1 \ldots l_{\kappa}}^{\dagger}}  
\right)^{\ominus} \gamma_{l_1 \ldots l_{\kappa}}. 
\end{eqnarray} 
Note that $U_g^{(\kappa)} (\mathcal{C}_{l_1},\ldots , \mathcal{C}_{l_{\kappa}})$ is a 
property of the ordered set of paths $\{ \mathcal{C}_{l_1},\ldots , \mathcal{C}_{l_{\kappa}} \}$ 
in the set of Grassmann manifolds $\{ \mathcal{G} (N;n_1), \ldots \mathcal{G} (N,n_{\kappa}) \}$. 
The GP in Eq. (\ref{eq:diagonal}) is contained as the special case where $\kappa = 1$. 

Similar to the  $\kappa = 1$ GPs discussed above, the phase factor 
$U_g^{(\kappa)} (\mathcal{C}_{l_1},\ldots , \mathcal{C}_{l_{\kappa}})$ is undefined or partially 
defined if $\gamma_{l_1 \ldots l_{\kappa}}$ vanishes or is not of full rank, respectively. These 
points are the singular points of the evolution of the system related to the off-diagonal GPs of 
order $\kappa$. It has been shown in Ref. \cite{kult07} that there is no singular point where 
all the different order non-Abelian GPs are undefined simultaneously.

\section{Realization of non-Abelian off-diagonal GPs in a four-qubit system}
Our four-qubit system is described by the Hamiltonian 
\begin{eqnarray}
\widetilde{H} =  F(s) \sum_{k=1}^4 \left( J_{k,k+1} R_{k,k+1}^{\textrm{XY}} + 
D_{k,k+1}^z R_{k,k+1}^{\textrm{DM}} \right) ,
\label{eq:4-qubit}
\end{eqnarray}
where $R_{k,k+1}^{\textrm{XY}} = \frac{1}{2} \left( \sigma_x^k \sigma_x^{k+1} + \sigma_y^k 
\sigma_y^{k+1} \right)$ and $R_{k,k+1}^{\textrm{DM}} = \frac{1}{2} \left( \sigma_x^k 
\sigma_y^{k+1} - \sigma_y^k \sigma_x^{k+1} \right)$ are XY and Dzialochinski-Moriya (DM)
terms with coupling strengths $J_{k,k+1}$ and $D_{k,k+1}^z$, respectively; $\sigma_x^k$ 
and $\sigma_y^k$ being standard Pauli operators acting on qubit $k$. $F(s)$ turns on and off 
all qubit interactions simultaneously. The cyclic nature of the qubit chain is reflected in the 
boundary conditions $J_{4,5} R_{4,5}^{\textrm{XY}} = J_{4,1} R_{4,1}^{\textrm{XY}}$ and 
$D_{4,5}^z R_{4,5}^{\textrm{DM}} = D_{4,1}^z R_{4,1}^{\textrm{DM}}$. 

The Hamiltonian in Eq. (\ref{eq:4-qubit}) preserves the single-excitation subspace 
\begin{eqnarray}
\mathcal{H}_{\text{eff}}=\textrm{Span} \{ \ket{1000},\ket{0010},\ket{0100},\ket{0001} \}
\label{eq:Heff}
\end{eqnarray}
of the four qubits. In the ordered orthonormal basis $\{ \ket{1000},\ket{0010},\ket{0100},
\ket{0001} \}$, the Hamiltonian takes the form 
\begin{eqnarray}
H =  F(s) \left( \begin{array}{rr}
0 & T \\
T^{\dagger} & 0 
\end{array} \right) ,
\label{eq:hamiltonian}
\end{eqnarray}
were  
\begin{eqnarray}
T = \left( \begin{array}{rr}
J_{12} - iD_{12}^z & J_{41} + iD_{41}^z  \\
J_{23} + iD_{23}^z  & J_{34} - iD_{34}^z  
\end{array} \right) = U S V^{\dagger} . 
\end{eqnarray}
Here, $U,V$, and $S$ are the unitary and diagonal positive parts in the singular-value 
decomposition of $T$. We assume $S>0$.   

The Hamiltonian in Eq. (\ref{eq:hamiltonian}) may be implemented in different physical 
systems. First, it may describe a cyclic chain of four coupled quantum dots, where the 
single excitation is encoded in the localized electron spins with double occupancy of 
each dot being prevented by strong Hubbard-repulsion terms \cite{mousolou12}. Secondly, 
a square optical lattice of two-level atoms with synthetic spin-orbit coupling localized at each 
lattice site allows for the desired combination of XY and DM interactions, by suitable parameter 
choices \cite{radic12}. A third possible realization is provided by the Ruderman-Kittel-Kasuya-Yosida 
interaction in three-dimensional topological insulators, which may be used to obtain the XY 
and DM interaction terms in $\widetilde{H}$ \cite{zhu11}.

The Hamiltonian in Eq. (\ref{eq:hamiltonian}) splits the effective state space into two orthogonal 
subspaces, i.e., 
\begin{eqnarray}
\mathcal{H}_{\text{eff}} = \mathcal{H}_1 \oplus \mathcal{H}_2,
\end{eqnarray}
where the subspaces
$\mathcal{H}_1$ and $\mathcal{H}_2$ are spanned by frames $\{ \ket{1000} , \ket{0010} \}$ 
and $\{ \ket{0100} , \ket{0001} \}$ respectively. This implies that the time evolution operator 
on the effective Hilbert space given in Eq. \ref{eq:Heff} splits into $2 \times 2$ blocks according 
to \cite{mousolou12}
\begin{eqnarray}
\mathcal{U} (t,0) = \left( \begin{array}{cc} 
U \cos \left( a_t S \right) U^{\dagger}  & 
-i U \sin \left( a_t S \right) V^{\dagger} \\ 
-i V \sin \left( a_t S \right) U^{\dagger} &  V \cos \left( a_t S \right) V^{\dagger}  
\end{array} \right),
\label{eq:na-u(t)}
\end{eqnarray} 
where $a_t = \int_0^t F(s) ds$ is the `pulse area'.

Considering paths $\mathcal{C}_1$ and $\mathcal{C}_2$ traversed by the two subspaces 
$\mathcal{H}_1$ and $\mathcal{H}_2$ under $\mathcal{U} (t,0)$, 
one may notice that these evolutions are purely geometric since the Hamiltonian vanishes 
along each of them separately. Thus, the four $2 \times 2$ blocks of 
the time evolution operator $\mathcal{U} (t,0)$ contains explicit information about the pair 
of paths $\mathcal{C}_1$ and $\mathcal{C}_2$ in the Grassmann manifold $\mathcal{G} (4;2)$ 
that can be fully captured by the non-Abelian off-diagonal GPs for $\kappa = 1$ and $2$. 
In fact, we find 
\begin{eqnarray}
\mathcal{U} (t,0) = \left( \begin{array}{cc} 
\sigma_{11}  & \sigma_{12} \\
\sigma_{21} & \sigma_{22}
\end{array} \right),
\end{eqnarray}  
from which we obtain  
\begin{eqnarray}
\sigma_{11} & = & U \cos \left( a_t S \right) U^{\dagger} , \ 
\sigma_{22} = V \cos \left( a_t S \right) V^{\dagger} ,
\nonumber \\ 
\gamma_{12} & = & \sigma_{12} \sigma_{21} =  -U \sin^2 \left( a_t S \right) U^{\dagger} , 
\nonumber \\ 
\gamma_{21} & = & \sigma_{21} \sigma_{12} = -V \sin^2 \left( a_t S \right) V^{\dagger} .  
\label{eq:quantities}
\end{eqnarray}
The $\kappa = 1$ and $\kappa = 2$ GPs can be found from these quantities as follows. 

By assuming that $\cos \left( a_t S \right)$ is full rank, we obtain the 
$\kappa=1$ GPs
\begin{eqnarray}
U_g^{(1)} (\mathcal{C}_1) & = & U \left| \cos \left( a_t S \right) \right|^{-1} 
\cos \left( a_t S \right) U^{\dagger} 
\nonumber \\ 
 & = & (-1)^c U Z^d U^{\dagger} , 
\nonumber \\ 
U_g^{(1)} (\mathcal{C}_2) & = & V \left| \cos \left( a_t S \right) \right|^{-1} 
\cos \left( a_t S \right) V^{\dagger} 
\nonumber \\ 
 & = & (-1)^c V Z^d V^{\dagger} , 
\end{eqnarray}
where $c,d = 0,1$ and $Z=\text{diag}\left\{1,-1\right\}$. These GPs are characterized by 
different sectors whose boundaries are given by pulse area values $a_t$ such that one or 
both eigenvalues of $\cos \left( a_t S \right)$ vanish. These points are singular points of 
the time evolution of the system, where the $\kappa = 1$ GPs are undefined or partially 
defined. Explicitly, when passing through a point where only one of the eigenvalues of 
$\cos \left( a_t S \right)$ vanishes, $d$ changes by one unit and the GPs switch abruptly 
as $(-1)^{c'} \hat{1} \leftrightarrow (-1)^c U Z U^{\dagger}$ and $(-1)^{c'} \hat{1} \leftrightarrow 
(-1)^c V Z V^{\dagger}$, where $\hat{1}$ is the $2 \times 2$ identity matrix and $c,c' = 0,1$. 
If both eigenvalues pass through zero simultaneously, only $c$ changes by one unit corresponding 
to an overall change of sign. Thus, in this case, the GPs switch abruptly as $U Z^d U^{\dagger} 
\leftrightarrow - U Z^d U^{\dagger}$ and $V Z^d V^{\dagger} \leftrightarrow - V Z^d V^{\dagger}$. 

To compute the off-diagonal $\kappa = 2$ GPs, we first note that $\sin^2 \left( a_t S \right) 
\geq 0$. Thus, in the case where both eigenvalues of $\sin \left( a_t S \right)$ are non-vanishing, 
we find the $\kappa = 2$ GPs
\begin{eqnarray}
U_g^{(2)} (\mathcal{C}_1,\mathcal{C}_2) = U_g^{(2)} (\mathcal{C}_2,\mathcal{C}_1) = -\hat{1} .
\end{eqnarray}
If one or both eigenvalues of $\sin \left( a_t S \right)$ vanish then the $\kappa = 2$ GPs are 
partially defined or undefined, respectively; these cases correspond to the $\kappa = 2$ 
singular points of the time evolution of the system. However, there is no abrupt switching 
associated with passage through any of these points since the $\kappa = 2$ GPs can 
only take the value $-\hat{1}$ when it is fully defined. Note that the $\kappa = 1$ and 
$\kappa = 2$ singular points are mutually exclusive since they are respectively associated 
with vanishing eigenvalues of $\cos \left( a_t S \right)$ and $\sin \left( a_t S \right)$. This 
confirms the result of Ref. \cite{kult07} that there are no points where all non-Abelian GPs 
are undefined.

The independence of the details of the paths $\mathcal{C}_1$ and $\mathcal{C}_2$ in 
the $\kappa = 2$ GPs is analogous to the single-qubit case, where the corresponding 
Abelian off-diagonal GP factors always take the value $-1$ except for cyclic  
evolution where it is undefined \cite{manini00,hasegawa01,hasegawa02}. 
However, in contrast, the non-Abelian case admits a richer off-diagonal GP structure due 
to the fact that different parallel transporting pulses do in general not commute. To see 
this, consider a pair of pulses where the first one is characterized by $\widetilde{T} = 
\widetilde{U} \widetilde{S} \widetilde{V}^{\dagger}$ such that $\sin \left( a_{\tilde{t}} 
\widetilde{S} \right) = Z$ or $\hat{1}$ (to assure parallel transport also during the 
second pulse), followed by an arbitrarily long second pulse
 characterized by $T = USV^{\dagger} \neq \widetilde{T}$. We may write 
the resulting time evolution operator after the second pulse as 
\begin{eqnarray}
\mathcal{U} (t,0) & = & \mathcal{U} (t,\tilde{t}) \mathcal{U} (\tilde{t},0) 
\nonumber \\ 
 & = & \left( \begin{array}{cc} 
\sigma_{12} \widetilde{\sigma}_{21} & \sigma_{11} \widetilde{\sigma}_{12}  \\
\sigma_{22} \widetilde{\sigma}_{21}  & \sigma_{21} \widetilde{\sigma}_{12} 
\end{array} \right) , 
\end{eqnarray}  
where $\sigma_{kl}$ are defined by $T$ and $a_t = \int_{\tilde{t}}^t F(s) ds$, while 
$\widetilde{\sigma}_{kl}$ are defined by $\widetilde{T}$ and $a_{\tilde{t}} = \int_0^{\tilde{t}} 
F(s) ds$. 

Thus, we obtain the $\kappa = 2$ GPs 
\begin{eqnarray}
U_g^{(2)} (\mathcal{C}_1,\mathcal{C}_2) & = & \left| \sigma_{11} \widetilde{\sigma}_{12} 
\sigma_{22} \widetilde{\sigma}_{21} \right|^{\ominus} \sigma_{11} \widetilde{\sigma}_{12} 
\sigma_{22} \widetilde{\sigma}_{21}, 
\nonumber \\ 
 U_g^{(2)} (\mathcal{C}_2,\mathcal{C}_1) & = & \left| \sigma_{22} \widetilde{\sigma}_{21} 
\sigma_{11} \widetilde{\sigma}_{12} \right|^{\ominus} \sigma_{22} \widetilde{\sigma}_{21} 
\sigma_{11} \widetilde{\sigma}_{12} .  
\end{eqnarray}
In the full rank case,  $U_g^{(2)} (\mathcal{C}_1,\mathcal{C}_2)$ and $U_g^{(2)} (\mathcal{C}_2,
\mathcal{C}_1)$ are unitaries different from $-\hat{1}$. In fact, if we consider pulses, where $\widetilde{\sigma}_{12}$ and $\widetilde{\sigma}_{21}$
commute with $\sigma_{11}$ and $\sigma_{22}$, then we find 
\begin{eqnarray}
U_g^{(2)} (\mathcal{C}_1,\mathcal{C}_2) & = & -\left| \sigma_{11}  
\sigma_{22} \right|^{-1} \sigma_{11}  
\sigma_{22} , 
\nonumber \\ 
 U_g^{(2)} (\mathcal{C}_2,\mathcal{C}_1) & = & -\left| \sigma_{22} 
\sigma_{11}  \right|^{-1} \sigma_{22}  
\sigma_{11}  , 
\label{eq:su-generators}
\end{eqnarray}
where we have used $\widetilde{\sigma}_{21} \widetilde{\sigma}_{12} = \widetilde{\sigma}_{12} \widetilde{\sigma}_{21} = -\hat{1}$. 
Therefore, from Eqs. (\ref{eq:su-generators}) and (\ref{eq:quantities}) it follows that the 
off-diagonal phases $U_g^{(2)} (\mathcal{C}_1,\mathcal{C}_2)$ and $U_g^{(2)} 
(\mathcal{C}_2,\mathcal{C}_1)$ could be any arbitrary SU(2) matrices for appropriate 
choices of $T$ and $\widetilde{T}$. For instance, the above conditions leading to Eq. 
(\ref{eq:su-generators}) can be met by a first pulse characterized by $\widetilde{T} = 
\lambda \hat{1}$ and $a_{\tilde{t}}=\frac{(2 m-1)\pi}{2 \lambda}$, $\lambda$ and $m$ 
being a real positive number and an integer, respectively; followed by a second pulse characterized 
by arbitrary $T$ and $a_{t}$ such that $\cos \left( a_t S \right)$ is full rank.  

\begin{figure}[h]
\centering
\includegraphics[scale=0.14]{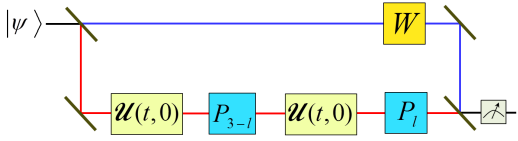}
\caption{Interferometric setting to measure the non-Abelian off-diagonal GP ($\kappa = 2$). The 
red and blue lines of the interferometer correspond to ancilla qubit states $\ket{0_a}$ and 
$\ket{1_a}$, respectively. An input state $\ket{0_a} \ket{\psi}$ with $\ket{\psi} \in \mathcal{H}_l, \ 
l=1,2$, enters into interferometer and splits into two equal-weighted state vectors each 
labeled by the ancilla basis states, i.e., $\frac{1}{\sqrt{2}}(\ket{0_a} + \ket{1_a})\ket{\psi}$.
This is achived by applying a Hadamard transformation to the input state of the ancilla qubit. 
Thereafter, the state attached to $\ket{0_a}$ undergoes successively the transformations 
$\mathcal{U} (t,0) $, $P_{3-l}$, and $\mathcal{U} (t,0)$, while $\ket{1_a}\ket{\psi}$ is left untouched. 
This is followed by performing the conditional transformation $\ket{0_a} \bra{0_a}\otimes P_l + 
\ket{1_a} \bra{1_a}\otimes W$. Next, the two state branches are brought back to interfere 
by a second Hadamard transformation. Finally, the probability $p$ of finding the final state at 
the output $\ket{0_a}$ branch is measured. By varying the unitary $W$,  maximum probability 
is obtain when $W = \Phi [\gamma_{l(3-l)}] = U_g^{(2)} (\mathcal{C}_l,\mathcal{C}_{3-l})$.}
\label{fig:interferometer}
\end{figure}

To test the non-Abelian off-diagonal GP ($\kappa = 2$), we add an ancilla qubit $\textrm{Span} 
\{ \ket{0_a},\ket{1_a} \}$ to the system, prepare the initial state in the superposition $\frac{1}{\sqrt{2}} 
\left( \ket{0_a} + \ket{1_a} \right)\ket{\psi}$, and perform conditional unitary dynamics 
\begin{eqnarray}
\frac{1}{\sqrt{2}} \left( \ket{0_a} + \ket{1_a} \right) \ket{\psi} \rightarrow 
\frac{1}{\sqrt{2}} \left( \ket{0_a} \mathcal{U} (t,0) \ket{\psi} + \ket{1_a} 
\ket{\psi} \right)  
\end{eqnarray}
with $\ket{\psi}$ belonging to $\mathcal{H}_l,\ l=1,2$. This transformation is followed by the 
operation $ \ket{0_a} \bra{0_a} \otimes P_{3-l}+ \ket{1_a} \bra{1_a}\otimes \hat{1}_a $, the 
conditional unitary above, and a operation $\ket{0_a} \bra{0_a}\otimes P_l +  
\ket{1_a} \bra{1_a}\otimes W $, where $P_l$ are projectors onto $\mathcal{H}_l$ and $W$ is a 
variable unitary onto $\mathcal{H}_l$. Finally, the ancilla states are transformed by a Hadamard, 
where $\ket{0_a} \rightarrow \frac{1}{\sqrt{2}} \left( \ket{0_a} + \ket{1_a} \right)$ and $\ket{1_a} 
\rightarrow \frac{1}{\sqrt{2}} \left( \ket{0_a} - \ket{1_a} \right)$. The resulting total state reads 
\begin{eqnarray}
\ket{\textrm{out}} & = & \frac{1}{2} \ket{0_a} \left( P_l \mathcal{U} (t,0) P_{3-l} \mathcal{U} (t,0) + 
W \right) \ket{\psi} 
\nonumber \\ 
 & & + \frac{1}{2} \ket{1_a} \left( P_l \mathcal{U} (t,0) P_{3-l}\mathcal{U} (t,0) - W \right) 
\ket{\psi} 
\end{eqnarray}
from which we read off the probability
\begin{eqnarray}
p & = & \parallel \ket{\textrm{out}} \parallel^2 = 
\frac{1}{4} + \frac{1}{4} \bra{\psi} \gamma_{l(3-l)} \gamma_{l(3-l)}^{\dagger} \ket{\psi} 
\nonumber \\ 
 & & + \frac{1}{2} \textrm{Re} \bra{\psi} W^{\dagger} \gamma_{l(3-l)} \ket{\psi} 
\end{eqnarray}
to detect the system in the state labeled by $\ket{0_a}$. By varying $W$ maximum probability 
is obtain when $W = \Phi [\gamma_{l(3-l)}] = U_g^{(2)} (\mathcal{C}_l,\mathcal{C}_{3-l})$. Thus, 
the off-diagonal GP can be measured by finding the maximum probability in the output of the 
interferometer depicted in Fig. \ref{fig:interferometer}. 

Alternatively, $U_g^{(2)} (\mathcal{C}_l,\mathcal{C}_{3-l})$ can be measured by realizing 
the interferometer loop directly on the input state $\ket{\psi} \in \mathcal{H}_l$ without 
adding the ancilla qubit. This results in the output state $W^{\dagger} P_l \mathcal{U} (t,0) 
P_{3-l} \mathcal{U} (t,0) \ket{\psi} = W^{\dagger} \gamma_{l(3-l)} \ket{\psi}$, which implies 
that the probability $\widetilde{p}$ to find the system in $\ket{\psi}$ satisfies 
\begin{eqnarray}
\widetilde{p} & = & \left| \bra{\psi} W^{\dagger} \gamma_{l(3-l)} \ket{\psi} \right|^2 
\nonumber \\ 
 & \leq & \left| \bra{\psi} \left( \sqrt{\gamma_{l(3-l)} \gamma_{l(3-l)}^{\dagger}} 
\right)^{\ominus} \ket{\psi} \right|^2
\end{eqnarray}
with equality when $W = \Phi [\gamma_{l(3-l)}] = U_g^{(2)} (\mathcal{C}_l,\mathcal{C}_{3-l})$ 
up to an overall U(1) phase factor. In this way, the non-Abelian SU(2) part of 
$U_g^{(2)} (\mathcal{C}_l,\mathcal{C}_{3-l})$ can be measured by varying $W$ until 
maximum is reached. 

We demonstrate how the latter setting can be implemented in the four-dot system 
mentioned above. As demonstrated in \cite{mousolou12}, a cyclic chain of coupled quantum 
dots at half-filling can be designed so that it is described by an effective spin Hamiltonian 
with XY and DM terms resulting from an interplay between electron-electron repulsion and 
spin-orbit interaction. With $\ket{\!\!\uparrow},\ket{\!\!\downarrow}$ being the local 
$s_z$-spin basis of each electron, the four-dimensional subspace $\ket{\!\!  \downarrow 
\uparrow\uparrow \uparrow}, \ldots, \ket{\! \! \uparrow\uparrow\uparrow \downarrow}$ 
is the invariant subspace $\mathcal{H}_{\text{eff}}$ in which the spin Hamiltonian takes the 
form of Eq. (\ref{eq:hamiltonian}) and where the XY and DM coupling strengths can be 
manipulated separately with time-dependent gate voltages. 

Now, to prepare an approriate initial state in the four-dot system, we start by polarizing the 
spins along the $z$ direction by an external magnetic field. A single spin flip is induced 
by applying a local magnetic field at one of the sites \cite{grinolds11}. Suppose, e.g., we 
apply it to the first site leading to the spin state $\ket{\psi} = \ket{\!\!  \downarrow \uparrow 
\uparrow \uparrow} \in \mathcal{H}_1 = \{ \ket{\!\!  \downarrow \uparrow\uparrow 
\uparrow}, \ket{\!\!  \uparrow \uparrow \downarrow \uparrow} \}$ of the four electrons. 
In this way, a measurement of $U(\mathcal{C}_1,\mathcal{C}_2)$ can be performed by 
applying sequentially $\mathcal{U} (t,0)$, $P_2 = \ket{\!\!  \uparrow \downarrow \uparrow 
\uparrow} \bra{\uparrow \downarrow \uparrow \uparrow \!\! } + \ket{\!\!  \uparrow 
\uparrow \uparrow \downarrow} \bra{\uparrow \uparrow \uparrow \downarrow \!\! }$, 
$\mathcal{U} (t,0)$, and $P_1 = \ket{\!\!  \downarrow \uparrow \uparrow \uparrow} 
\bra{\downarrow \uparrow \uparrow \uparrow ~\!\!} + \ket{\!\!  \uparrow \uparrow 
\downarrow \uparrow} \bra{\uparrow \uparrow \downarrow \uparrow \!\!}$, where 
$\mathcal{U} (t,0) = e^{-ia_t H}$, followed by the unitary $W^{\dagger} = e^{ib_{t'} h}$. 
The final unitary should be block diagonal with respect to the two orthogonal spin 
subspaces $\mathcal{H}_1$ and $\mathcal{H}_2$, which is achieved by implementing    
\begin{eqnarray}
h & = & f(s) \sum_{k=1}^2 \left( J_{k,k+2} R_{k,k+2}^{\textrm{XY}} + 
D_{k,k+2}^z R_{k,k+2}^{\textrm{DM}} \right) 
\nonumber \\ 
 & & + E (Z_1 + Z_2) . 
\end{eqnarray}
Here, $b_{t'} = \int_0^{t'} f(s) ds$ is the `pulse area' and $E (Z_1 + Z_2)$ with $Z_1 = 
\ket{\!\!  \downarrow \uparrow \uparrow \uparrow} \bra{\downarrow \uparrow \uparrow 
\uparrow \!\!} - \ket{\!\!  \uparrow \uparrow \downarrow \uparrow} \bra{\uparrow \uparrow 
\downarrow \uparrow \!\!}$ and $Z_2 = \ket{\!\!  \uparrow \downarrow 
\uparrow \uparrow} \bra{\uparrow \downarrow \uparrow \uparrow \!\! } - 
\ket{\!\!  \uparrow \uparrow \uparrow \downarrow} \bra{\uparrow \uparrow 
\uparrow \downarrow \!\! }$ corresponds to a local energy shift of the first and second 
sites relative the third and fourth site (for instance by applying an inhomogeneous 
magnetic field over the four-dot system). In the single spin flip subspace, $h = {\textrm{diag}} 
\{ T',T''\}$ with the $2\times 2$ blocks
\begin{eqnarray}
T' & = & \left( \begin{array}{cc}
E & J_{13} + iD_{13}^z  \\
J_{13} - iD_{13}^z  & -E 
\end{array} \right) , 
\nonumber \\ 
T'' & = & \left( \begin{array}{cc}
E & J_{24} + iD_{24}^z  \\
J_{24} - iD_{24}^z  & -E 
\end{array} \right). 
\end{eqnarray}
The variable unitary $W^{\dagger}$ is generated by $h$ and takes the desired block-diagonal form 
$W^{\dagger} = {\textrm{diag}} \{ e^{ib_{t'} T'},e^{ib_{t'} T''}\}$ with $e^{ib_{t'} T'}$ and $e^{ib_{t'} T''}$
being arbitrary SU(2) operators parametrized by $J_{k,k+1},D_{k,k+1}$, and $E$. Thus, with inital 
state $\ket{\psi} \in \mathcal{H}_1$, the $\kappa = 2$ GP $U(\mathcal{C}_1,\mathcal{C}_2)$ can 
be measured by varying the parameters $J_{13}, D_{13},$ and $E$ until the probability $\widetilde{p}$ 
reaches its maximum. 

\section{Conclusions}
In conclusion, we have demonstrated a setup which admits direct observation of the non-Abelian
off-diagonal geometric phases (GPs). The system consists of four qubits arranged in a cyclic chain 
and nearest-neighbor interaction of combined XY and Dzialoshinski-Moriya type. We have shown that
the off-diagonal GPs span the full SU(2) group by applying sequentially different pulsed interactions 
between the qubits. The resulting off-diagonal GPs can be observed in an interferometric setting.  
\vskip 0.3 cm
C.M.C. and V.A.M were supported by Department of Physics and Electrical Engineering at 
Linnaeus University (Sweden) and by the National Research Foundation (VR). E.S. acknowledges 
support from the National Research Foundation and the Ministry of Education (Singapore).


\begin{thebibliography}{99}
\bibitem{manini00} MANINI N. and PISTOLESI F.,
{\it Phys. Rev. Lett.}, {\bf 85} (2000) 3067.
\bibitem{hasegawa01} HASEGAWA Y., LOIDL R., BARON M., BADUREK G. 
and RAUCH H.,
{\it Phys Rev. Lett.}, {\bf 87} (2001) 070401. 
\bibitem{hasegawa02} HASEGAWA Y., LOIDL R., BADUREK G., BARON M., 
MANINI N., PISTOLESI F. and RAUCH H.,
{\it Phys Rev. A}, {\bf 65} (2002) 052111.
\bibitem{filipp03a} FILIPP S. and SJ\"OQVIST E.,
{\it Phys. Rev. Lett.}, {\bf 90} (2003) 050403 (2003). 
\bibitem{filipp03b} FILIPP S. and SJ\"OQVIST E.,
{\it Phys. Rev. A}, {\bf 68} (2003) 042112. 
\bibitem{tong05} TONG D. M., SJ\"OQVIST E., FILIPP S., 
KWEK L. C. and OH C. H.,
{\it Phys. Rev. A}, {\bf 71} (2005) 032106.
\bibitem{kult07} KULT D., {\AA}BERG J. and SJ\"OQVIST E.,
{\it EPL.}, {\bf 78} (2007) 60004. 
\bibitem{bengtsson06} BENGTSSON I. and \.{Z}YCZKOWSKI K., 
{\it Geometry of quantum states} (Cambridge University Press,
Cambridge) 2006, Ch. 4.9.
\bibitem{anandan89} ANANDAN J. and PINES A., 
{\it Phys. Lett. A} {\bf 141}, 335 (1989).  
\bibitem{sjoqvist06}  SJ\"OQVIST E., KULT D. and {\AA}BERG J.,
{\it Phys. Rev. A}, {\bf 74} (2006) 062101.
\bibitem{zanardi99} ZANARDI P. and RASETTI M.,
{\it Phys. Lett. A}, {\bf 264} (1999) 94. 
\bibitem{sjoqvist12} SJ\"OQVIST E, TONG D. M., ANDERSSON L. M., HESSMO B., 
JOHANSSON M. and SINGH K., 
{\it New J. Phys.}, {\bf 14}, 103035 (2012). 
\bibitem{mousolou12} MOUSOLOU V. A., CANALI C. M. and SJ\"OQVIST E., 
arxiv:1209.3645.
\bibitem{radic12} J. RADI\'{C} J., DI CIOLO A., K. SUN K. and V. GALITSKI V., 
{\it Phys. Rev. Lett.} {\bf 109} (2012) 085303. 
\bibitem{zhu11} ZHU J.-J., YAO D.-X., ZHANG S.-C. and CHANG K., 
{\it Phys. Rev. Lett.} {\bf 106} (2011) 097201. 
\bibitem{wilczek84} WILCZEK F. and ZEE A.,
{\it Phys. Rev. Lett.}, {\bf 52} (1984) 2111.
\bibitem{moore20} MOORE E. H.,
{\it Bull. Am. Math. Soc.}, {\bf 26} (1920) 394. 
\bibitem{penrose55} PENROSE R.,
{\it Proc. Cambridge Phil. Soc.}, {\bf 51} (1955) 406.
\bibitem{mostafazadeh99} MOSTAFAZADEH A.,
{\it J. Phys. A}, {\bf 32} (1999) 8157.
\bibitem{kult06} KULT D., {\AA}BERG J. and SJ\"OQVIST E.,
{\it Phys. Rev. A}, {\bf 74} (2006) 022106.
\bibitem{grinolds11} GRINOLDS M. S., MALETINSKY P., HONG S., LUKIN M. D., 
WALSWORTH R. L. and YACOBY A., 
{\it Nature Phys.}, {\bf 7} (2011) 687. 	
\end{thebibliography}
\end{document}